\documentclass[3p,times,procedia]{elsarticle}
\usepackage{nupha_ecrc}

\volume{00}

\firstpage{1}

\journalname{Nuclear Physics A}

\runauth{Wilke van der Schee}

\jid{nupha}

\jnltitlelogo{Nuclear Physics A}

\usepackage{amssymb}

\usepackage[figuresright]{rotating}
\usepackage{color}
\usepackage{graphicx}

\usepackage{siunitx}
\begin{document}

\begin{frontmatter}



\dochead{XXVIth International Conference on Ultrarelativistic Nucleus-Nucleus Collisions\\ (Quark Matter 2017)}

\title{Equilibration and hydrodynamics at strong and weak coupling}


\author[label1,label2]{Wilke van der Schee}			

\address[label1]{Center for Theoretical Physics, MIT, Cambridge MA 02139, USA}
\address[label2]{Institute for Theoretical Physics and Center for Extreme Matter and Emergent Phenomena, Utrecht University, Leuvenlaan 4, 3584 CE Utrecht, the Netherlands}

\begin{abstract}
We give an updated overview of both weak and strong coupling methods
to describe the approach to a plasma described by viscous hydrodynamics,
a process now called hydrodynamisation. At weak coupling the very
first moments after a heavy ion collision is described by the colour-glass
condensate framework, but quickly thereafter the mean free path is
long enough for kinetic theory to become applicable. Recent simulations
indicate thermalization in a time $t\sim40(\eta/s)^{4/3}/T$ \cite{Keegan:2015avk},
with $T$ the temperature at that time and $\eta/s$ the shear viscosity
divided by the entropy density. At (infinitely) strong coupling it
is possible to mimic heavy ion collisions by using holography, which
leads to a dual description of colliding gravitational shock waves.
The plasma formed hydrodynamises within a time of $0.41/T$. A recent
extension found corrections to this result for finite values of the
coupling, when $\eta/s$ is bigger than the canonical value of $1/4\pi$,
which leads to $t\sim(0.41+1.6(\eta/s-1/4\pi))/T$ \cite{Grozdanov:2016zjj}.
Future improvements include the inclusion of the effects of the running
coupling constant in QCD.
\end{abstract}


\begin{keyword}
	hydrodynamisation \sep holography


\end{keyword}

\end{frontmatter}



\section{Introduction}

The creation of a strongly coupled plasma at relativistic nucleus-nucleus
collisions is one of the most striking discoveries at RHIC and LHC.
One of the hallmarks resulting from this program is the understanding
that this plasma is described by viscous relativistic hydrodynamics
very quickly, within 1 fm/c. This is surprising since the gradients
at that time result in a small longitudinal pressure, even with the
small viscosity present. This process of going to a regime described
by hydrodynamics is now called hydrodynamisation and depending on
the gradients present this can take much shorter than the process
of equilibration.

It is profoundly challenging to describe the entire process from collision
to hydrodynamics fully within QCD itself. This is partly because at
high energy scales a perturbative treatment can be appropriate, while
at energy scales of the temperature of the plasma formed a strong
coupling picture should be used. 

A perturbative treatment is not straightforward, since at higher energies
the gluon concentration in nuclei increases. After the collision it
is hence natural to expect an overoccupied state of gluons, which
act coherently in a state called colour-glass condensate (CGC). At
weak coupling the CGC undergoes classical expansion, up to the point
where the mean-free path is long enough to allow for a description
using kinetic theory. At strong coupling there are currently no theoretical
tools to do such a description within QCD itself, although within holography it is possible
to try and mimic QCD as close as possible \cite{Gursoy:2007cb}.

This talk will review both weak and strong coupling approaches from
a far-from-equilibrium initial stage to a plasma described by relativistic
hydrodynamics. This includes both the timescale of the process, as well
as the evolution of the flow and energy density during the process.

\section{Hydrodynamisation at weak and strong coupling}

At weak coupling the first stage of a heavy ion collision is described
by classical Yang-Mills \cite{Gelis:2013rba,Berges:2013eia}. After a short time
the mean free path of the gluons becomes long enough that evolution
using kinetic theory becomes feasible. For pure Yang-Mills this was
first achieved in \cite{Kurkela:2015qoa}, where it was found that
for 't Hooft coupling $\lambda=10$ a typical state hydrodynamises
within a time of $\tau\sim10/Q_{s}$, with $Q_{s}$ the saturation
scale. In terms of the temperature at hydrodynamisation and the shear
viscosity this translates into $\tau\sim40(\eta/s)^{4/3}/T$ \cite{Keegan:2015avk}.
More recently, this has been extended to also include dynamics in
the transverse plane \cite{Keegan:2016cpi}, which is useful since
among else it can shed light on the early time dynamics of radial
flow. At early times this pre-flow in the transverse plane grows linearly
with time and for any approximately boost-invariant conformal theory
is given by \cite{Vredevoogd:2008id}
\begin{equation}
v_{\perp}=-\frac{\tau}{2}\frac{\nabla_{\perp}e}{e+P_{\perp}},
\end{equation}
where $e$ is the energy density profile, and $P_{\perp}$ the transverse
pressure. For accurate initial hydrodynamic conditions the relevant
question is then what the (average) transverse pressure is during
the far-from-equilibrium evolution. For strong coupling the transverse
pressure starts out high, at $P_{\perp}=e$, but decreases fast, giving
an average effective pressure of approximately $P_{\perp}\approx e/2$
\cite{vanderSchee:2013pia,Habich:2014jna}. In \cite{Keegan:2016cpi}
it is instead found that at weak coupling the transverse pressure
starts at $e/2$ and does not change much at early times. In the end
both weak and strong coupling approaches hence give rise to very similar
transverse flow.

\begin{figure}
\begin{centering}
\includegraphics[width=8.5cm]{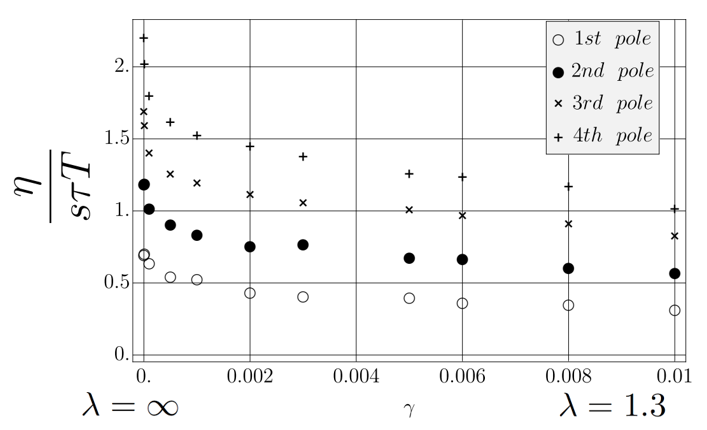}\,\,\,\includegraphics[width=5.5cm]{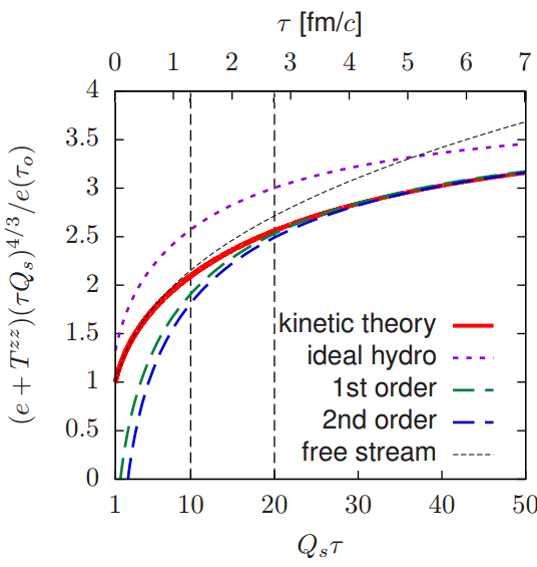}
\par\end{centering}

\caption{Left we show relaxation times $\tau$ as a function of the 't Hooft coupling $\lambda$ for a partially resummed supersymmetric Yang-Mills theory, 
in units of the temperature $T$ and the viscosity over entropy ratio $\eta/s$ (from \cite{Grozdanov:2016vgg}, see also \cite{Waeber:2015oka}). Even though the $\eta/s$ ranges from 0.08 to 1.05 in the range plotted, the ratio is remarkably constant. Right we show the relaxation of the energy density for a particular evolution using gluonic kinetic theory (from \cite{Kurkela:2016mhu}), for $\lambda=10$ (i.e. $\eta/s\approx 0.6$ \cite{Arnold:2000dr}). It can be seen that the system is well described by (viscous) hydrodynamics within approximately 2 fm/c.
\label{fig:relaxation}}
\end{figure}

For hydrodynamisation at strong coupling there have been studies in
a homogeneous \cite{Chesler:2008hg,Heller:2012km} and boost-invariant
setting \cite{Chesler:2009cy,Heller:2011ju}, finding a far-from-equilibrium
regime almost immediately followed by a near-equilibrium regime described
by quasi-normal modes. These systems were always well described by
hydrodynamics within a time of $1/T$, with $T$ the temperature at
that time of hydrodynamisation, which is non-trivial for the varying
energy density in the boost-invariant case.

Recently, there have been elaborate studies studying similar dynamics
in theories of higher derivative gravity \cite{Waeber:2015oka,Grozdanov:2016vgg}
(see also \cite{Stricker:2013lma}). These are especially interesting
for heavy ion collision, as these higher derivative terms can correspond
to inverse coupling constant corrections on the field theory side,
and can hence give insights into a coupling constant regime more akin
to QCD. In these papers higher derivative corrections to $\mathcal{N}=4$
SYM are considered, as well as Einstein-Gauss-Bonnet theory. Corrections
to $\mathcal{N}=4$ SYM itself are perhaps more interesting from a
fundamental point of view, since the field theory is known precisely.
Nevertheless, this higher derivative theory does not contain curvature
squared ($R^{2})$ or curvature cubed terms due to the symmetry of
this particular theory. For general higher derivative corrections
to more general QFTs it is hence more logical to consider a general
approach, starting to look at curvature squared theories, which is
given by the Einstein-Gauss-Bonnet action\footnote{Any theory with 2nd order curvature corrections can be written as
Einstein-Gauss-Bonnet theory to leading order in $\lambda_{GB}$ by
performing suitable redefinitions \cite{Brigante:2007nu,Grozdanov:2014kva}, which is
particularly nice, since it makes the equations of motion 2nd order.}:
\begin{equation}
S_{GB}=\frac{1}{2\kappa_{5}^{2}}\int d^{5}x\sqrt{-g}\biggr[R-2\Lambda+\frac{\lambda_{GB}}{2}\left(R^{2}-4R_{\mu\nu}R^{\mu\nu}+R_{\mu\nu\rho\sigma}R^{\mu\nu\rho\sigma}\right)\biggr].
\end{equation}
The works \cite{Waeber:2015oka,Grozdanov:2016vgg} resolved a puzzle
found in \cite{Stricker:2013lma}, where it was found that higher
order quasi-normal mode frequencies (QNM, determining relaxation time)
received much larger corrections than lower order QNMs. This would
quickly make them the dominant QNM, with very long relaxation times.
In the examples studied in \cite{Waeber:2015oka,Grozdanov:2016vgg},
however, it was found that this effect is really only there when treating
$\lambda_{GB}$ perturbatively. When using realistic values, and computing
the relaxation time non-perturbatively, the QNMs remain ordered, albeit
giving longer relaxation times: $\tau  \sim \frac{0.72}{2 \pi T}(1+(\eta/s-1/4\pi)7.7)$,
for the case of $\mathcal{N}=4$ SYM \cite{Waeber:2015oka} (Figure \ref{fig:relaxation}, left).
See also the recent paper \cite{Grozdanov:2016fkt} for a complete analysis of hydrodynamics
and its breakdown in theory dual to Einstein-Gauss-Bonnet.

\begin{figure}
\begin{centering}
\includegraphics[width=8.0cm]{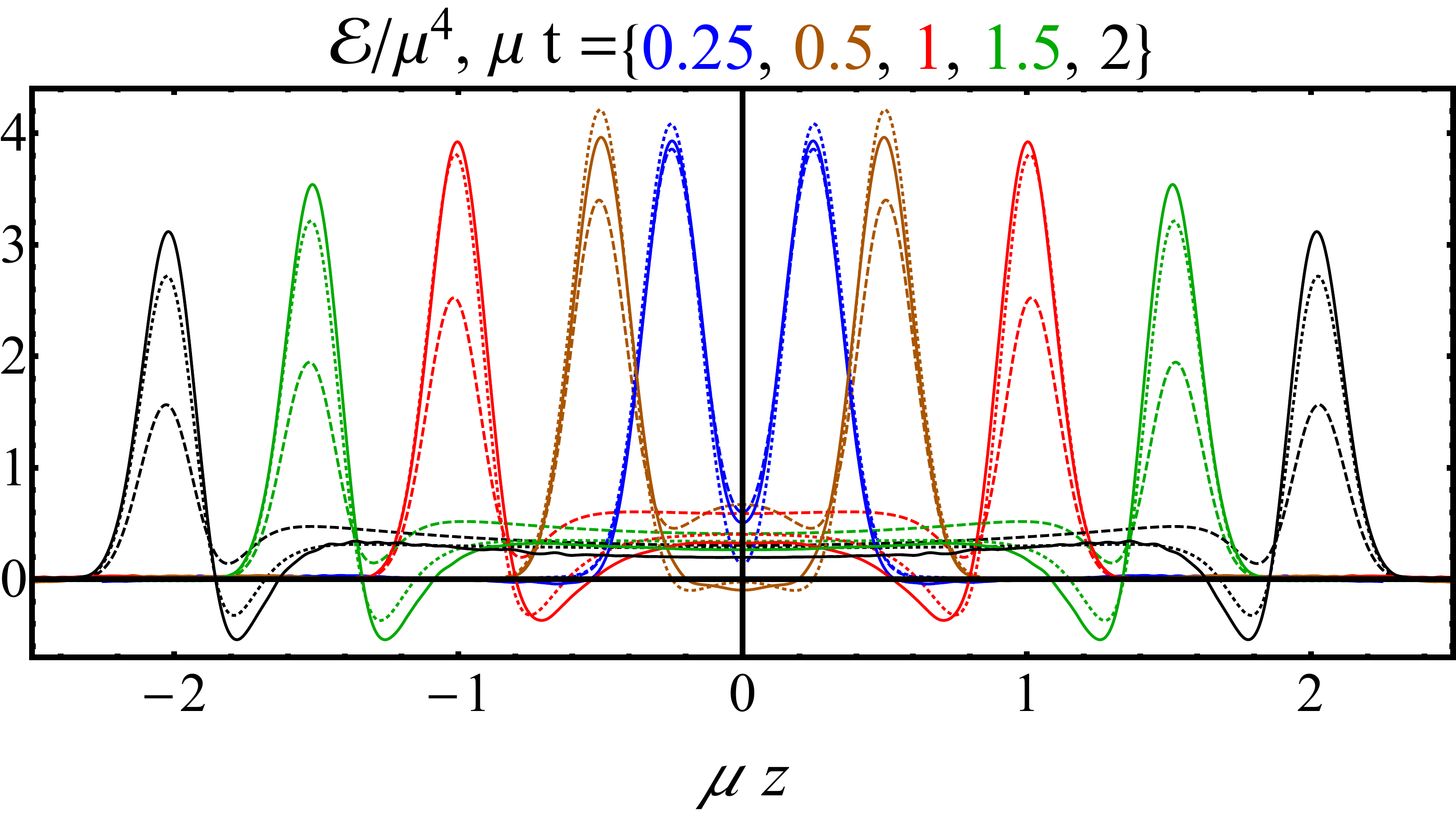}\,\,\,\, \includegraphics[width=6.3cm]{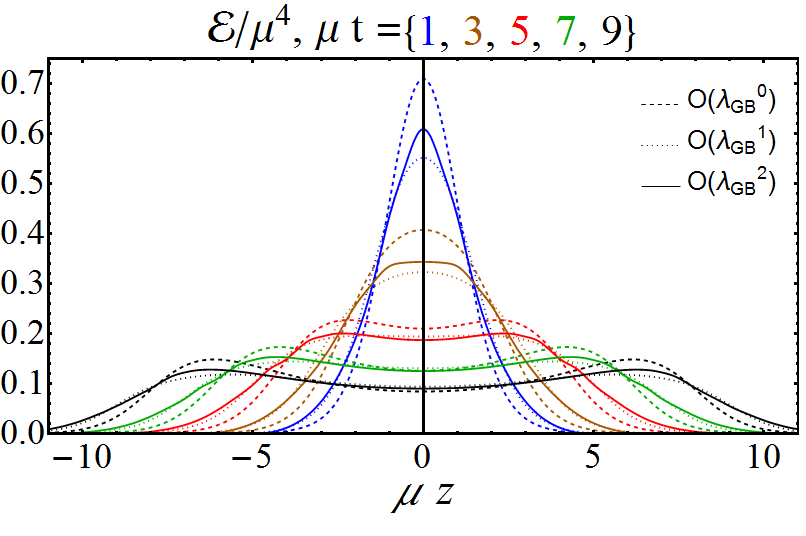}
\par\end{centering}

\caption{We show the energy density in units of the impact energy density for
narrow (left) and wide (right) holographic heavy ion collisions for
both infinitely strong coupling (dashed), and including finite coupling
corrections (solid for 2nd order corrected, dotted for 1st order and dashed for uncorrected). The simluations have $\lambda_{GB}=-0.2$, corresponding
to $\eta/s=1.8/4\pi$, which would give about $\lambda = 8.0$ in the case of $\mathcal{N}=4$ SYM theory (figure from \cite{Grozdanov:2016zjj}).\label{fig:GBshocks}}
\end{figure}

More realistic for heavy ion collisions is to set up an actual collision,
homogeneous in the transverse plane, whereby the initial conditions
are given blobs of plasma boosted to the speed of light, while keeping
their total energy and longitudinal profile fixed \cite{Chesler:2010bi,Casalderrey-Solana:2013aba,Casalderrey-Solana:2013sxa}.
Since the simplest versions of holography are scale invariant conformal
field theories a stationary blob of plasma would start expanding,
but the time dilatation from boosting to the speed of light makes
them stable (see however \cite{Attems:2017zam} for collisions in
non-conformal theories). For these `holographic heavy ion collisions'
it was found that when the longitudinal profile is thin enough (`narrow'
shocks) all physics only depends on the energy per transverse area,
which we call $\mu^{3}$. In this case the plasma is described by
hydrodynamics very quickly, within a time of $t\,T\sim0.41$, depending slightly
on the criterion how accurately hydrodynamics should describe the evolution
of the system (we use 10\% accuracy for the pressures).

Recently, these collisions have been extended to Einstein-Gauss-Bonnet
gravity, treated perturbatively in $\lambda_{GB}$, up to 2nd order. Snapshots of the
energy density for both narrow shocks (left, Gaussian with width $\mu w=0.1$)
and wide shocks (right, $\mu w=1.5$) can be seen in Figure \ref{fig:GBshocks}.
The results clearly illustrate that for the narrow shocks (high energy) there is
more energy on the lightcone and less energy in the plasma, indicating that indeed
interactions are weaker and that the collision is hence more transparent.
For the wide shocks two effects are visible. Firstly, also here less energy ends up
at mid-rapidity, there is less 'pile-up'. Secondly, the energy is flatter, which can also 
be seen in the narrow case. The latter effect can be explained by the larger viscosity
which in this case leads to a larger anisotropy and hence smaller longitudinal pressure, which
in turns leads to less acceleration in the longitudinal profile.
For these collisions the corrected hydrodynamisation time turns out to be
$t\sim(0.41+1.6(\eta/s-1/4\pi))/T$. For wider shocks the time is a bit more involved,
since then the width of the shocks provides a more important scale than the temperature
and a more elaborate analysis needs to be done.

\begin{figure}
\begin{centering}
\includegraphics[width=7.5cm]{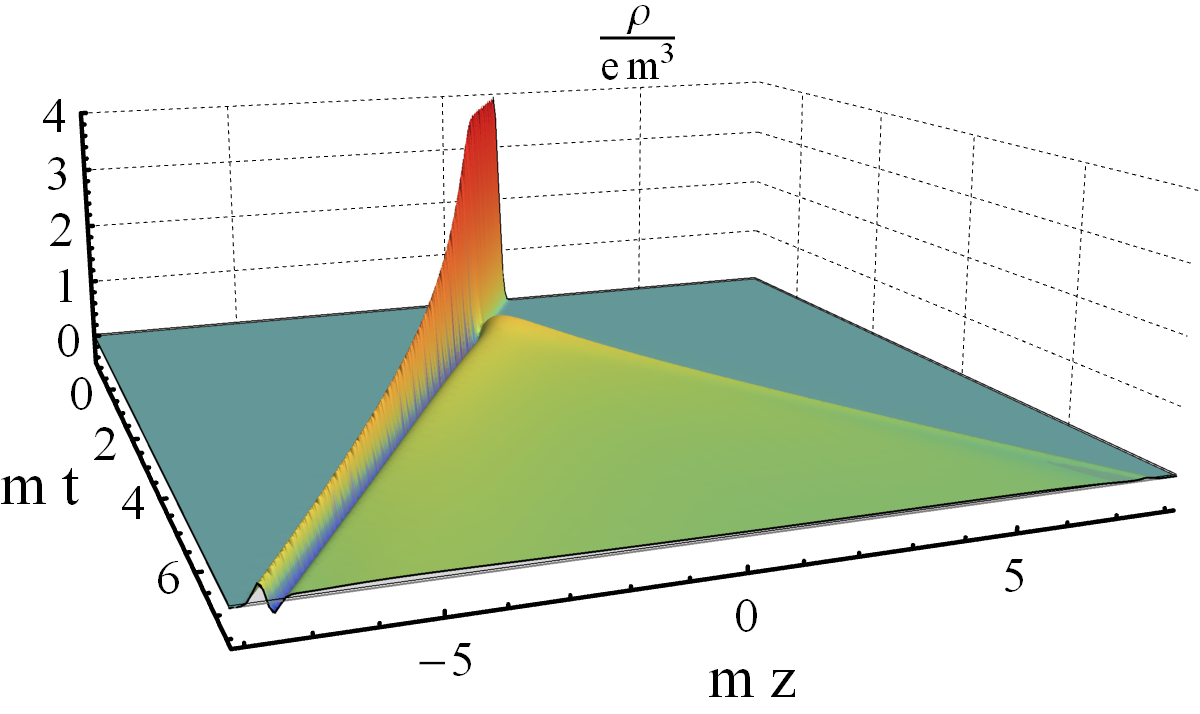}\,\,\quad \includegraphics[width=5cm]{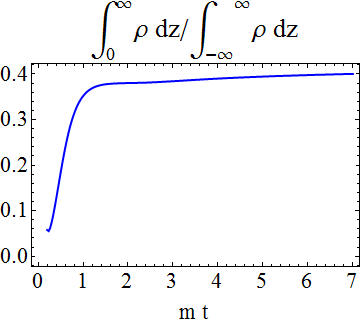}
\par\end{centering}

\caption{We present the charge density $\rho$ for a collision where only the
left-moving shock is charged as a function of time and longitudinal
direction (left) and with the fraction of the charge ending up at
$z>0$ as a function of time (right) ($m$ is equal to $\mu$ in Fig.
\ref{fig:GBshocks}, to avoid confusion with the chemical potential).
Interestingly, even though the initial charge moves towards negative
$z$ at the speed of light the collision causes about 41\% ends up
at positive $z$, indicating very strong interactions (figure from \cite{Casalderrey-Solana:2016xfq}).\label{fig:chargedshocks}}
\end{figure}

Another interesting study in the context of holograhpic heavy ion
collisions was done in \cite{Casalderrey-Solana:2016xfq}, which studied
collisions including a globally conserved U(1) charge. It was found
that the charge behaves much like the energy density, inclduing similar
(charged) hydrodynamisation times, as well as similar rapidity profiles\footnote{Quite curiously, the decay of the charge on the lightcone exactly
mimics the decay of the energy density. While perhaps natural from
a field theory perspective, this is surprising in terms of the gravitational
theory, where energy density is determined by non-linear dynamics
of curved spacetime and charge evolves according to linear Maxwell
equations on top of this geometry.}. Including charge, however, gives rise to an interesting opportunity
to collide equal energy shocks whereby only one shock is charged,
leaving the other shock neutral. This asymmetric set up allows a clean
probe of which part of the charge in the plasma comes from which shock
(this is true even in the symmetric case, as in our small chemical
potential approximation the charge-charge collision is just the sum
of two charge-neutral collisions).

Quite strikingly, Fig. \ref{fig:chargedshocks} shows that after a
short time more than 41\% of the right-moving charge is in fact moving
left, i.e. 41\% of the charge `bounced' off the other shock, as seen
in the centre-of-mass frame. This is to be contrasted with a weak
coupling picture, where all charge would just move through the shock,
perhaps loosing some speed, but not easily changing direction. This
hence shows an interesting qualitative difference in the two pictures.
From an experimental point of view this 'bounce' is probably not realistic,
at least not at top RHIC energies, where indeed most baryon charge
is found near the incoming nuclei \cite{Bearden:2003hx}.

Apart from differences in hydrodynamisation times interesting qualitative
differences arise at very early times directly after the collision.
Conformal invariance and energy conservation implies that in a boost-invariant
setting the stress-tensor in $(\tau,\,\eta,\, \bold{x}_{\perp})$
coordinates is given by
\begin{equation}
T_{\mu}^{\nu}={\rm diag}\left\{ -\epsilon(\tau),\,-\epsilon(\tau)-\tau\,\epsilon'(\tau),\,\epsilon(\tau)+\frac{1}{2}\tau\,\epsilon'(\tau),\,\epsilon(\tau)+\frac{1}{2}\tau\,\epsilon'(\tau)\right\} ,
\end{equation}
with $\epsilon(\tau)$ the energy density and $\tau$ proper time.
At early times there are three distinct interesting possibilities,
each leading to different pressures:
\begin{enumerate}
\item Free streaming of weakly interacting particles gives $\epsilon(\tau)\sim1/\tau$,
which is dominated by the expanding volume. This leads to $\mathcal{P}_{L}=T_{\eta}^{\eta}=0$.
\item In CGC fields act coherently and one has $\epsilon(\tau)\sim const+\mathcal{O}(\tau)$,
whereby the fields act cohorently. This can lead to a negative longitudinal
pressure: $\mathcal{P}_{L}=-\epsilon(\tau)$.
\item At strong coupling the plasma takes time to form, and one has $\epsilon(\tau)\sim\tau^{2}+\mathcal{O}(\tau^{5})$
\cite{Grumiller:2008va}. This leads to $\mathcal{P}_{L}=-3\epsilon(\tau)$.
\end{enumerate}
Each scenario hence has different qualitative initial time behaviour,
especially when focussing on the longitudinal pressure. Nevertheless,
for the hydrodynamic initial conditions they may lead to similar conclusions,
since at stronger coupling the pressure starts out more negative,
but also increases faster.

\section{Rapidity profile}

The study of equilibration and hydrodynamization does not only give
information about the time scale of the process. From the simulations
it is perhaps even more important to extract the hydrodynamic initial
data, which is given by the velocity and temperature fields (or equivalently
the energy density in the local restframe). At weak coupling the rapidity
profile is studied with the glasma framework in \cite{Schenke:2016ksl}
(see also proceedings in this issue, as well as \cite{Ipp:2017lho}).
Even though the glasma does not hydrodynamise by itself, this paper
is able to also find an approximately Gaussian rapidity profile, with
a width proportional to the inverse coupling $1/\alpha_{s}$.

\begin{figure}
\begin{centering}
\includegraphics[width=7.5cm]{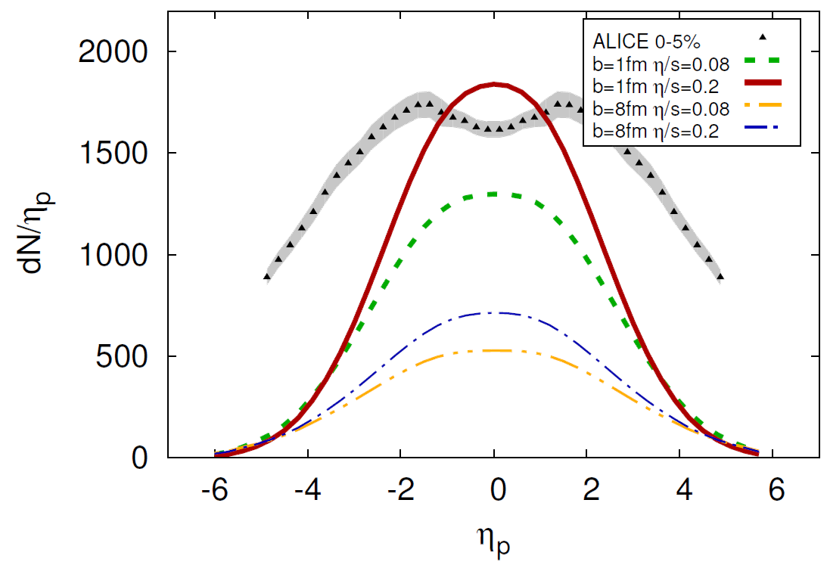}\,\,\,\includegraphics[width=7cm]{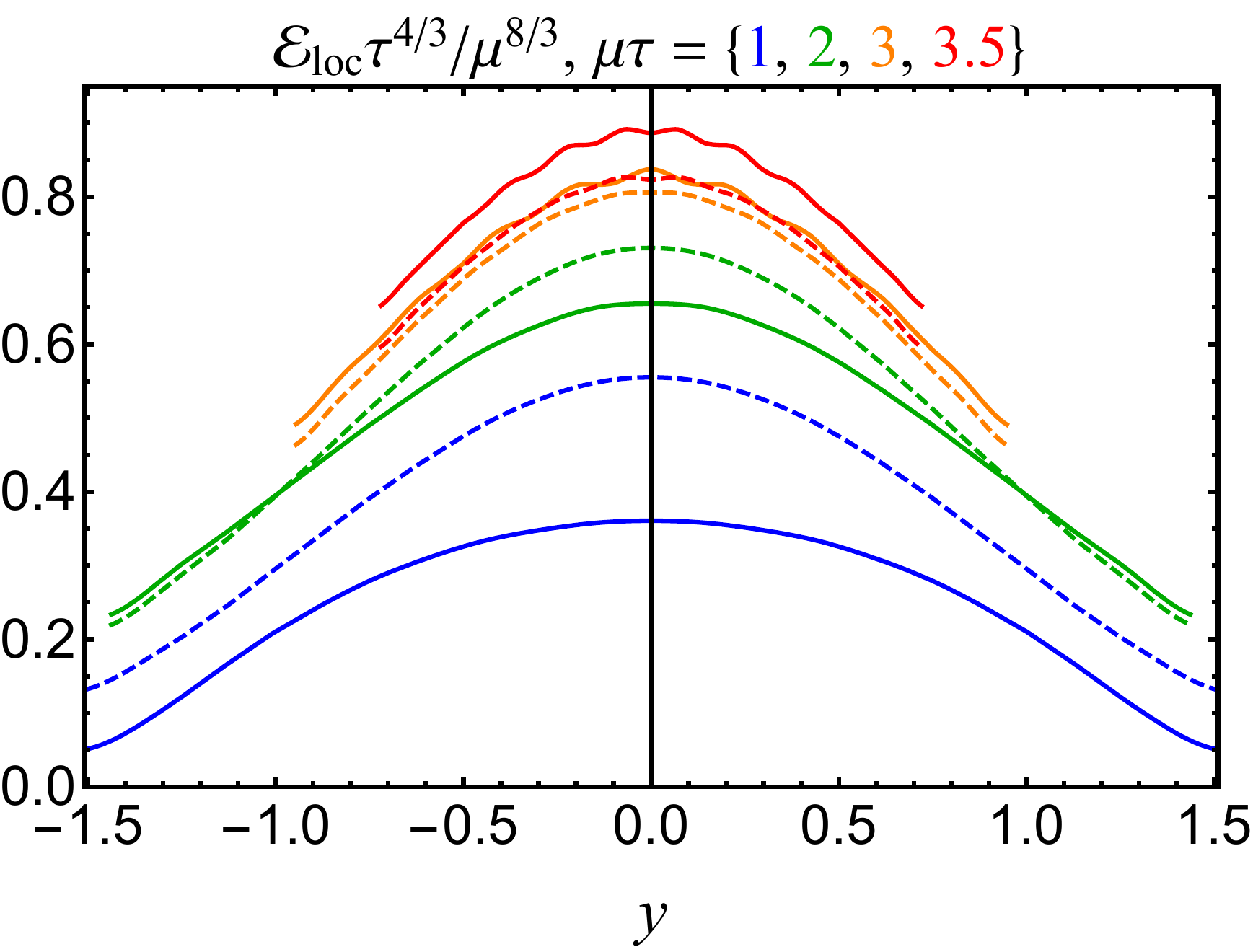}
\par\end{centering}

\caption{The rapidity profile for the infinitely coupled shocks in Figure \ref{fig:GBshocks}
turns out (from \cite{vanderSchee:2015rta,Grozdanov:2016zjj})\label{fig:rapidity-profile}}
\end{figure}

At strong coupling it was also found that the rapidity profile is Gaussian
to a very good approximation, with width 0.955 \cite{Chesler:2015fpa}.
This, however, is the initial rapidity profile at the moment of hydrodynamisation.
Hydrodynamic evolution will widen this profile, which was studied
by the MUSIC hydrodynamic and freeze-out code in \cite{vanderSchee:2015rta}.
This allows to directly compare with the experimental pseudo-rapidity
distribution as measured by ALICE (Figure \ref{fig:rapidity-profile},
left). This pseudo-rapidity profile is cleary wider than the Gaussian
of width 0.955, but also significantly narrower than the experimental
result. This calculation was a the full result of a completely collision in the simplest
holographic setting, at infinitely strong coupling, with no fitting
parameters are available. 

At this point one has to realise that this computation is not performed
in QCD itself, so for a realistic computation one either has to introduce
phenomenological fitting parameters (such as promoting the width 0.955
to a parameter), or to consider more advanced versions of holography
that are closer to QCD itself. For this latter option the attempt presented
in Figure \ref{fig:GBshocks} would be such an example. As explained 
these shocks in Einstein-Gauss-Bonnet gravity have a higher shear
viscosity and can in many ways be viewed as being described by  a weaker
coupling \cite{Grozdanov:2016vgg}. The rapidity profile is hence
of crucial interest and it presented in Figure \ref{fig:rapidity-profile}
(right). Indeed at early times the energy density is smaller, since
more energy remains on the lightcone due to the weaker coupling. At
the same time the rapidity profile is slightly wider, as clearly seen
for $\mu\tau=1\text{ and }2$. At later times, however, the energy
density becomes bigger than the unperturbed theory, and the width
of the rapidity profile does not increase as fast as the unperturbed
shock collisions. Both features can be understood by realising that this
theory has a higher shear viscosity. This leads to higher viscous entropy production,
which in turn leads to higher energy densities in the local restframe. Also, with
the gradients present the higher viscosity will lead to a larger anisotropy and hence
a smaller longitudinal pressure, which explains why in hydrodynamic evolution the 
rapidity profile does not widen as much as in the canonical collision.
Of course these collisions are performed using a single value of $\eta/s$, whereas
in QCD the viscosity is expected to vary with the local temperature. In real heavy
ion collisions the magnitude of both effects would depend on this variation of the shear viscosity.

\section{Discussion}



The initial stage of heavy ion collisions progresses from weaker coupling
at the moment of the collision to strong coupling in the later hydrodynamic
regime. In this talk I have reviewed the two main approaches of hydrodynamisation,
both currently working at constant coupling. The first approach employs
kinetic theory and is applicable at weak coupling, whereas the holographic
approach is valid at strong coupling for theories that share similarities
with QCD.

A somewhat separate question we discussed is the initial condition
at the moment the system hydrodynamised. This is expected to both
depend on the initial condition for the far-from-equilibrium evolution,
as well as the on the evolution itself. At weak coupling the initial
condition for the kinetic theory would ideally come from a first principle
computation using classical Yang-Mills theory \cite{Gelis:2013rba,Berges:2013eia}
extended with a rapidity analysis in glasma such as in \cite{Schenke:2016ksl}.
At strong coupling a natural initial condition is the collision of
lumps of energy. These always consist of gravitational waves in the
dual gravity theory, but they can be supplemented by vector (for a
conserved charge, \cite{Casalderrey-Solana:2016xfq}), scalar (for
a non-conformal theory \cite{Attems:2017zam}) or higher order gravity
corrections (for finite coupling corrections \cite{Grozdanov:2016zjj}).

In future it would be greatly beneficial to have simulations that
have a coupling constant evolving with the energy scale. Treating
the entire evolution before hydrodynamisation at weak coupling will
likely not be accurate at later times, where would be strongly coupled
with small viscosity (even at $\lambda=10$ in kinetic theory the
viscosity is still large, with $\eta/s\approx0.6$ at about $\tau=1$
fm/c). Treating the entire evolution at strong coupling, on the other
hand, will likely lead to too strong interactions at very early times,
which is illustrated by all the energy and charge ending up around
mid-rapidity (Figures \ref{fig:chargedshocks} and \ref{fig:rapidity-profile}).
A combination of both approaches may however lead to a realistic simulation
throughout the evolution.

\section*{Acknowledgments}

It is a pleasure to thank Jorge Casalderrey-Solana, Sa\v{s}o Grozdanov, Michal Heller, Aleksi Kurkela, David Mateos, Paul Romatschke, Bj\"{o}rn Schenke and Miquel Triana
for collaboration on projects presented here. This work is supported
by the U.S. Department of Energy under grant Contract Number DE-SC0011090.





\bibliographystyle{elsarticle-num}
\bibliography{references}






\end{document}